\documentclass[aps,prl,reprint,floatfix,superscriptaddress,showpacs,amsfonts
s,amssymb,amsmath,preprintnumbers]{revtex4-1}
\usepackage{bm}
\usepackage{epsfig}
\usepackage{amsmath}
\usepackage{amssymb}
\usepackage{graphicx}
\usepackage{bm}
\usepackage{color}
\usepackage{subfigure}

\newcommand{\fig}[1]{Fig.~\ref{#1}}

\newcommand{\eq}[1]{Eq.~(\ref{#1})}
\newcommand{\eqs}[2]{Eqs.~(\ref{#1}--\ref{#2})}

\newcommand{\be}{\begin{equation}}
\newcommand{\ee}{\end{equation}}

\newcommand{\bem}{\begin{multline}}
\newcommand{\eem}{\end{multline}}

\newcommand\bea{\begin{eqnarray}}
\newcommand\eea{\end{eqnarray}}

\newcommand{\curl}{\bm \nabla \times}

\begin{document}
\setcounter{secnumdepth}{1}

\title{The fully kinetic Biermann battery and associated generation of pressure anisotropy}

\author{K. M. Schoeffler}
\affiliation{GoLP/Instituto de Plasmas e Fus\~ao Nuclear,
Instituto Superior T\'ecnico,\\ Universidade de Lisboa, 1049-001 Lisboa, Portugal}
\author{N. F. Loureiro}
\affiliation{Plasma Science and Fusion Center, Massachusetts Institute of Technology, Cambridge MA 02139, USA}
\author{L. O. Silva}
\affiliation{GoLP/Instituto de Plasmas e Fus\~ao Nuclear,
Instituto Superior T\'ecnico,\\ Universidade de Lisboa, 1049-001 Lisboa, Portugal}

\date{\today}

\begin{abstract}

The dynamical evolution of a fully kinetic, collisionless system with imposed
background density and temperature gradients is investigated analytically. The
temperature gradient leads to the generation of temperature anisotropy, with
the temperature along the gradient becoming larger than that in the direction
perpendicular to it. This causes the system to become unstable to pressure
anisotropy driven instabilities, dominantly to electron Weibel.  When both
density and temperature gradients are present and non-parallel to each other,
we obtain a Biermann-like linear in time magnetic field growth.  Accompanying
particle in cell numerical simulations are shown to confirm our analytical
results.

\end{abstract}


\maketitle


\paragraph{Introduction.}

Both the seed field required for the generation of astrophysical magnetic
fields \cite{Kulsrud92,Kulsrud08} and intense magnetic fields generated in laser-solid
interaction laboratory experiments~\cite{Stamper71,Li07,Gregori12,Gao15} have
been attributed to the Biermann battery~\cite{Biermann50}. The Biermann battery
mechanism generates magnetic fields due to non-parallel temperature and
density gradients. Until now, the understanding of this mechanism has been
restricted to fluid models where an extra non-ideal term is added to Ohm's law.
In weakly or non-magnetized plasmas, the validity of fluid models rests on
collisions being sufficiently frequent compared to the dynamic timescales of
the problem, such that the pressure tensor remains in scalar form \footnote{In
systems with large magnetic fields such that the Larmor radius is small
compared to the system size (not relevant in this work) fluid models with a
non-scalar pressure tensor aligned with the field can be formulated
\cite{Chew56}.}.  These conditions are often not present in astrophysical
environments and are questionable in some laser-plasma environments, and thus a
fully kinetic model is necessary.

Recently, the Biermann battery has been investigated with fully self-consistent
kinetic 3D simulations~\cite{Schoeffler14, Schoeffler16}, but a clear
theoretical model for how the fully kinetic Biermann battery actually works in
collisionless plasmas has not been presented.  Such a model is presented here
for the first time, explaining not only the kinetic Biermann battery but also,
more generally, the dynamical evolution of collisionless unmagnetized plasmas
subject to background density and temperature gradients. In addition to
extending the validity of the Biermann battery to many weakly collisional
scenarios, we reveal the purely kinetic result that a temperature gradient
alone leads to the generation of anisotropies in temperature (pressure tensor).
This anisotropy gives rise to kinetic instabilities such as the Weibel
instability~\cite{Weibel59}, seen in~\cite{Schoeffler14, Schoeffler16}, or
instabilities that inhibit the heat flux~\cite{Levinson92,Gary00} on time
scales short compared to the collision time. This is relevant for a wide
variety of settings including astrophysical shocks and laser experiments with
small collision rates, and aids in the understanding of cooling flows in galaxy
clusters, which cannot be explained assuming the larger heat flux predicted
based solely on collisions~\cite{Fabian94}.

\paragraph{Model.}
We solve the time evolution of the velocity distribution function and
electromagnetic fields according to the coupled Vlasov and Maxwell's equations,
assuming that only the electrons play a role and the ions are static, only
acting as a neutralizing background.  For our calculation, we normalize the
velocity $\bm{v}$ to $v_{T0}\equiv \sqrt{T_e/m_e}$, time $t$ to
$\omega_{pe}^{-1}$, and $\bm{x}$ to $\lambda_D$, where $\omega_{pe}$ is the
plasma frequency for density $n = n_0$, and $\lambda_D \equiv
v_{T0}/\omega_{pe}$ is the Debye length. In addition $E$ and $B$ are normalized
to $E_0 \equiv m_e v_{T0} \omega_{pe}/e$ and $B_0 \equiv m_e c \omega_{pe}/e$
respectively. 

We will assume that a background Maxwellian distribution function, $f_M$, is
instantaneously perturbed such that
\be
n = n_0\left(1 + \epsilon x \right)\text{, } v_{T} = v_{T0} \sqrt{1 + \delta y}
\text{,}
\ee
\be
\epsilon \equiv \frac{\lambda_D}{L_n} \equiv \lambda_D\frac{1}{n}\frac{\partial n}{\partial x}\left(0\right),\text{ }
\delta \equiv \frac{\lambda_D}{L_T} \equiv \lambda_D \frac{1}{T}\frac{\partial T}{\partial y}\left(0\right)
\text{.}
\ee

This perturbation is not an equilibrium solution; it will be taken as a given
initial state.  Note that the Biermann battery is not an instability (in fluid
models it grows linearly, not exponentially, with time; we will find here that
this remains true in the kinetic case). As such, it has to arise from a
non-equilibrium state.

The parameters $\epsilon$ and $\delta$ are taken to be very small and
comparable to each other; they will be used as our asymptotic expansion
coefficients.  Assuming $\bm x \sim \epsilon^0$, the initial distribution
function to second order in $\epsilon$ and $\delta$ is:
\begin{multline}
\label{f0}
f_0 = f_M +
\epsilon x f_M
- \frac{1}{2} \delta y \left(3-v^2\right) f_M\\
+ \frac{1}{8} \delta^2 y^2 \left(15 -10v^2 +v^4 \right) f_M
-  \frac{1}{2} \epsilon\delta xy \left(3-v^2\right) f_M 
\text{.}
\end{multline}
We evolve the Vlasov-Maxwell equations initialized with this distribution
function, and either no initial electric or magnetic fields, or natural
equilibrium fields that act to balance the force due to the pressure gradient.

The evolution of the electron distribution function subject to these density
and temperature gradients is given by the Vlasov equation, coupled with
Faraday's and Ampere's laws: 
\begin{equation}
\label{vlasov}
\frac{\partial f}{\partial t} + \mathbf{v} \cdot \bm \nabla f 
- \left( \mathbf{E}
+ \mathbf{v}Â \times \mathbf{B} \right)
\cdot \bm \nabla_{v} f = 0
\text{,}
\end{equation}
\begin{equation}
\label{faraday}
\frac{\partial \mathbf{B}}{\partial t} = -\curl \mathbf{E}
\text{,}
\end{equation}
\begin{equation}
\label{ampere}
\frac{\partial \mathbf{E}}{\partial t} = \int d^3v \mathbf{
v}f + \frac{c^2}{v_{T0}^2} \curl \mathbf{B}
\text{.}
\end{equation}

We will seek solutions to these equations in powers of $\epsilon$ and $\delta$.
We will assume $t \sim  \bm{x} \sim c^2/v_{T0}^2 \sim
\epsilon^0 \sim \delta^0$.  Although the solution is only valid when $\bm{x}
\sim \epsilon^0$, at an arbitrary position $\bm{x}$, the calculation remains
valid in a new coordinate system $\bm{x}^{\prime }$, where the assumptions are
satisfied using $\epsilon^\prime$ calculated with the local $v_{T0}^\prime$ and
$n_0^\prime$. There are three other small parameters besides $\epsilon$ and
$\delta$; namely $c_s/v_{T0}$, $v_{T0}^2/c^2$, and $\nu/\omega_{pe}$, where
$c_s$ is the sound speed, and $\nu$ is the collision frequency. Each of these
parameters are assumed to be much smaller than one, but aside from
$\nu/\omega_{pe}$, can in principle remain of order $\epsilon^0$. \footnote{Our
calculation assumes $v_{T0}^2/c^2 \sim \epsilon^0$, although as long as $\bm B
\sim \epsilon^2$ (as we find in our solution), it is acceptable for
$v_{T0}^2/c^2 \sim \epsilon^1$.} We implicitly assume small values for these
parameters by assuming static ions, using the non-relativistic Vlasov equation/
Maxwellian distribution, and ignoring collisions. 

First we highlight some important aspects of the form of the solution.
The first order ($\sim \epsilon^1$) solution including all terms proportional
to $c_s/v_{T0}$ and $v_{T0}^2/c^2$, of $\bm E$, and $f$ is uniform in space,
and $f$ is an odd function of $v$. A proof of this is provided in the
supplementary materials. Given a uniform $E$, from \eq{faraday} no magnetic field
is generated, and an odd $f$ with respect to $v$ only leads to uniform bulk
flows and temperature fluxes. It is thus necessary that we perform our
calculation with second order terms ($\sim \epsilon^2$) to see the Biermann
battery, and the formation of a temperature anisotropy.  The second order
solution is different in form, and except for terms of $f$ which are even in
$v$, there are no terms that are uniform in space. It should be emphasized that
this means that modifications coming from $c_s/v_{T0}$ and $v_{T0}^2/c^2$ can
be separately neglected for both first order and second order solutions. Note
that second order modifications to the first order solution are then neglected
(the entire solution is not accurate to $\epsilon^2$).

Solutions can be found from an initial condition by taking an expansion for
small $t$, restricted to second order in $\epsilon$. Fortunately,
the sum over all orders of $t$ converges to a solution valid for arbitrary
$t \sim \epsilon^0$, and thus only small compared to the electron transit
time $L_T/v_{T0} = \delta^{-1}$. 

\paragraph{Density gradient.}
We first consider the case with only a density gradient ($\delta=0$). If
we assume the initial condition of $f = f_0$ and no initial electric or
magnetic fields, we obtain the following analytic solution:
\begin{equation}
\label{perturbedfdensity}
f = f_0 + \tilde{f_n}\left(t\right)
\text{,}
\end{equation}
\begin{equation}
\label{perturbedEdensity}
\bm{E} 
= -
\left(\epsilon
-\epsilon^2
x \right)
\left[1-\cos\left(\omega_{pe,x}t\right)\right]
\hat{\bm x}
\text{,}
\end{equation}
where $\omega_{pe,x} \equiv 1+\epsilon x/2$ is the normalized plasma frequency
based on the $x$ dependent density, $n$, and $\tilde{f_n}$ is an oscillatory
term described in the supplementary materials.  It is evident that the electric
field of this solution oscillates about
\be
\label{equilibriumEdensity}
\bm{E} = -
\left(\epsilon
-\epsilon^2
x  \right)
\hat{\bm x}
\text{.}
\ee

The space dependent frequency, $\omega_{pe,x}$, gives rise to increasingly
shorter scale $x$ variations of the electric field. These variations along $x$
lead to phase mixing in space and then Landau damping. Our model does not show
this damping because the damping is exponentially supressed until $k\lambda_D
\lesssim 1$ which occurs at $t \sim \epsilon^{-1}$ where the assumptions break
down.  Eventually Landau damping eliminates the oscillations, and thus the
electric field should naturally settle to \eq{equilibriumEdensity}. If we take
\eq{equilibriumEdensity} as the initial condition for the electric field, we
arrive at an equilibrium solution to \eqs{vlasov}{ampere} where $\bm{E}$ and
$f$ do not change with time. 

\paragraph{Temperature gradient.}
We now consider a second case, with a temperature gradient only 
($\epsilon=0$, $\delta\ne0$).  If we again start with the initial conditions
$f = f_0$, and no initial electric or magnetic fields, to second order in
$\delta$, the solution to \eqs{vlasov}{ampere} is the following:
\begin{equation}
\label{perturbedfpressure}
f =
f_{\nabla T} + \tilde{f_T}\left(t\right) 
\text{,}
\end{equation}
\be
\label{perturbedEpressure}
\bm{E}  
= 
-\delta
\left[1-\cos\left(t \right)\right]
\hat{\bm y}
\text{,}
\ee
where
\begin{multline}
\label{fTevolution}
f_{\nabla T} \equiv f_0 +
\frac{1}{2}\delta  t v_y\left(5 -v^2\right) f_M\\
-\frac{1}{4}\delta^2  t y v_y \left(25 - 12v^2 + v^4 \right) f_M\\
+\delta^2  t^2  
\left[ \frac{1}{8} v_y^2 \left(39 - 14v^2 + v^4\right) - \frac{1}{4}\left( 5- v^2\right)\right] f_M
\text{,}
\end{multline}
and $\tilde{f_T}$ is an oscillatory term described in the supplementary materials.
Once again the solution oscillates around a particular value for the electric field:
\be
\label{equilibriumEpressure}
\bm{E} 
= - \delta
\hat{\bm y}
\text{.}
\ee
Although Landau damping at the gradient scale with a wavelength $k \lambda_D
\sim \delta$ is exponentially supressed, in a system with density gradients,
this field would also Landau damp as described earlier. Therefore it is
similarly natural to consider starting from \eq{equilibriumEpressure} as the
initial condition.  This yields a simpler solution where the electric field is
constant with time, but the distribution function continues to evolve with time
as $f = f_{\nabla T}$. 

Two important terms in \eq{fTevolution} grow with $t$ and eventually break the
assumptions of the ordering. The second term on the RHS of \eq{fTevolution} is
associated with the heat flux, and matches the collisional solution shown
in~\cite{Levinson92} once $t$ reaches the collision time. However the
assumptions will have already broken when $t \sim \delta^{-1}$.  The fourth term on the RHS
of \eq{fTevolution}, which grows as $t^2$ is associated with a temperature
anisotropy, where the collisionless temperatures in each direction
(corresponding to diagonal components of the pressure tensor) differ.  This
term breaks the assumptions earlier; when $t \sim \delta^{-1/2}$ (a hybrid
between the plasma period and electron transit time $\left( L_T/v_{T0}
\omega_{pe} \right)^{1/2}$). However, the simulations will show that the
predictions remain valid beyond this limit.

We define the pressure tensor (normalized to $m_e n_0 v_{T0}^2$) as:
\be
\label{temperaturedef}
n T_{ij} \equiv \int d^3v v_iv_j f 
\text{,}\\
\ee
from which we find
$T_{yy} = v_T^2 + 3/2 \delta^2 t^2$ and 
$T_{xx} = T_{zz} = v_T^2 + 1/2 \delta^2 t^2$,
resulting in the following anisotropy:
\begin{equation}
\label{anisotropydef}
A \equiv \frac{T_{yy}}{T_{xx}}-1 = \delta^2 t^2
\text{.}
\end{equation} The temperature gradient thus naturally leads to a temperature
anisotropy. More energetic particles up the gradient arrive faster if their
momentum is predominantly along the gradient. This anisotropy will give rise to
kinetic instabilities such as the Weibel instability~\cite{Weibel59} seen
in~\cite{Schoeffler14} or instabilities that inhibit the heat
flux~\cite{Levinson92,Gary00}. 

\begin{figure}
  \noindent\includegraphics[width=3.5in]{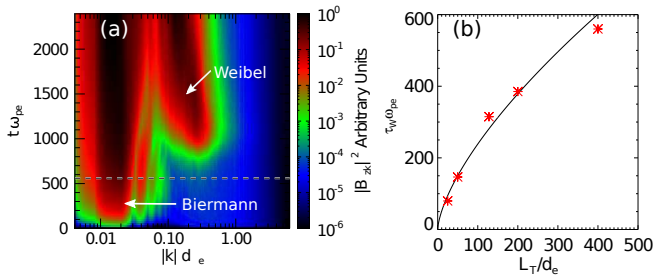}
  \caption{\label{Weibelgrowthtimes}
	  (a) Magnetic energy spectra of $B_z$ (with respect to $|k|=\sqrt{k_x^2+k_y^2}$) vs. time from the simulation with system size
  $L_T/d_e=400$ $(L_T/\lambda_D= 2000)$ reported in Ref.~\cite{Schoeffler14}.
  The time when the Weibel instability begins to grow exponentially is identified with a dashed line. This
  estimate of the onset time $\tau_W$ is plotted vs. system size (b) along with the predicted curve where \eq{weibelonset} is satisfied 
  indicating where the Weibel growth rate exceeds that of the anisotropy predicted in \eq{anisotropydef}.
}
\end{figure}

\eq{anisotropydef} is consistent with the anisotropy and the subsequent
development of the Weibel instability obtained in the PIC simulations reported
in \cite{Schoeffler14} --- see ~\fig{Weibelgrowthtimes}.  The onset time of the
Weibel instability $\tau_W$ is roughly estimated from the magnetic energy spectra
when the Weibel field begins to grow exponentially. ~\fig{Weibelgrowthtimes}(a)
shows the spectra for the case where $L_T/d_e = 400$ $(\delta^{-1} = 2000)$, with the onset of Weibel indicated by a dashed line.
Although the Biermann field is energetically dominant for smaller system sizes,
the higher $k$ Weibel instability is present (i.e. an onset time can be
measured) for all simulations. The onset time, $\tau_W$,  should occur when the Weibel
growth rate, which is a function of anisotropy, and thus of time, exceeds the predicted rate of anisotropy growth from
\eq{anisotropydef}:
\begin{equation}
\label{weibelonset}
\gamma_W(A(\delta t)) > 1/A \partial A /\partial t = 2/t
\text{.}
\end{equation}
\fig{Weibelgrowthtimes}(b) shows that this prediction matches the estimated
onset over a range of system sizes remarkably well, where $\gamma_W(A)$ is the
growth rate of the Weibel instability, given by \cite{Weibel59} , which we
solve numerically. $\gamma_W$ is calculated at the location of fastest growth
($x/L_T=0.9125, y/L_T=0$), which is independent of system size using the local
values; $v_{T} = 0.036 c$, $n= 0.12 n_0$, and the anisotropy calculated from
\eq{anisotropydef} using $L_{T,local} = 0.0625 L_T$ and $v_T$.  Note that this
anisotropy is slightly increased by a factor of $5/4$ to take into account
second order variations in temperature, which will be addressed in a future
work. \footnote{A second derivative of the temperature, which is greater along
$\hat{x}$, leads to a temperature anisotropy hotter along $\hat{x}$. ($A = t^2 (\partial^2 T /\partial x^2 - \partial^2 T /\partial y^2)/T $)}

It is surprising that the agreement is so good since these simulations are in
highly nonlinear regimes; the assumption that $\tau_W \ll \delta^{-1}$ is only
satisfied for sufficiently large $L_T$/$d_e$.  For the largest $L_T/d_e = 400$
case ($\tau_W \omega_{pe,local} \approx 0.8 \delta_{local}^{-1}$), nonlinear
effects were clearly present. The thermal velocity, which we assume to be
constant with time except for the small modification $\sim \delta^2t^2$, grew as
$v_T \sim t$. The measured anisotropy grew at close to $A \sim t^4$, which is
still consistent with $A = \delta^2t^2$, given that $\delta$ is now a function
of time. We expect the onset time to continue to follow this trend for even
larger $L_T$, where our assumption $\tau_W \omega_{pe,local} \ll
\delta_{local}^{-1}$ is valid.

\paragraph{Biermann battery.}
Both of these simplified cases begin oscillating about the equilibrium electric
fields, \eq{equilibriumEdensity} and \eq{equilibriumEpressure}, at the
timescale of the electron plasma frequency, and Landau damping eventually
eliminates these oscillations. These equilibrium fields balance the associated
pressure gradients.  It is thus a natural assumption to start with a similar
electric field for the initial conditions for the complete case
($\epsilon\ne0$, $\delta\ne0$): 
\be
\bm{E} =  
\label{Ebiermann}
 -
\left(\epsilon
-\epsilon^2
x 
+ \epsilon\delta
y 
 \right)
\hat{\bm x}
-
\delta 
\bm{\hat{y}}
\text{.}
\ee
With this assumption, and starting with $f_0$,
the solution to \eqs{vlasov}{ampere} is:
\begin{equation}
\label{fbiermann}
f = f_{\nabla T}
+
\frac{1}{2} \epsilon\delta  
t 
x v_y \left(5 -v^2\right) f_M
+ \frac{1}{2}\epsilon\delta  
t^2 
v_x v_y f_M
\text{,}
\end{equation}
with the magnetic field:
\be
\label{Bbiermann}
\bm{B}
= -\epsilon\delta
t 
\hat{\bm z}
\text{,}
\ee
where the electric field does not change with time.

The last term of~\eq{fbiermann} is associated with the off-diagonal component
of the pressure tensor, and thus $T_{xy}$, which enhances the magnitude and rotates the
direction of the temperature anisotropy (defined in the frame
that diagonalizes $T_{ij}$). 

We thus see a fully kinetic Biermann battery: the magnetic field grows linearly
in time, and is proportional to both the density and temperature gradients, as
in the fluid case.

\paragraph{Numerical comparison.}
Our analytic model has been tested via particle-in-cell (PIC) simulations
using the OSIRIS framework  \cite{Fonseca02, Fonseca08}.  The simulations are
done setting  $\epsilon = \delta = 0.001$, and a normalized thermal velocity
$v_{T0}/c = 0.05$, which is small such that relativistic effects do not play a
role, but large compared to $\epsilon$.  A more detailed explanation of the
simulation parameters and setup is outlined in the supplementary materials

To test these solutions we look at both simulations with $\delta=0$ or
$\epsilon=0$, and with both gradients. Good agreement between the predicted and
simulated electric fields for single gradients is shown in the supplementary
materials.
\begin{figure}
  \noindent\includegraphics[width=3.0in]{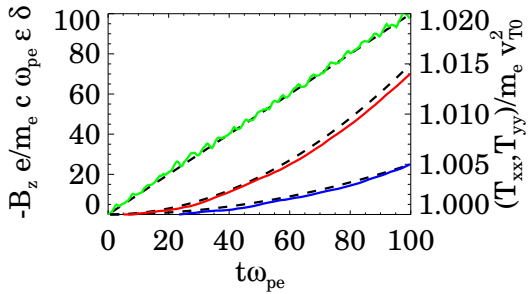}
  \caption{\label{kineticBiermannplot}
  Evolution of the averaged $B_z$ due to the perpendicular
  density and temperature gradients (green), and anisotropic temperatures $T_{xx}$ (blue) and $T_{yy}$
  (red), vs. time along with the predicted curves at $y=0$ from \eq{Bbiermann} and
  \eq{temperaturedef} (black dashed lines). 
}
\end{figure}
In~\fig{kineticBiermannplot} the average magnetic field from the simulation
with both gradients grows linearly in time, confirming the prediction in
\eq{Bbiermann}.  The growth of the temperatures in the $x$ and $y$ directions
shown in~\fig{kineticBiermannplot} matches quite well with the predictions from
\eq{temperaturedef}. Note that our solution is rigorously only valid for $t \ll
\delta^{-1/2}=50$, but this breaking would occur at many more
$\omega_{pe}^{-1}$ for realistically small values of $\delta$ that are not
feasible to simulate.  The plotted simulation fields are calculated by
averaging the results between $-49 < x < 49$, and $-49 < y < 49$.  

\paragraph{Conclusions.}
In this Letter, we have presented analytical solutions of the Vlasov-Maxwell
system of equations for collisionless systems with background density and
temperature gradients. The kinetic equivalent of the Biermann battery --- a
linear in time magnetic field growth --- has been obtained for the first time.
Another noteworthy result is the generation of temperature anisotropy in all
cases where a background temperature gradient is present. This implies that
pressure anisotropy driven instabilities, such as electron Weibel, should be
expected in such systems. These may have a profound impact on their evolution,
from effectively determining the magnetic field growth, to constraining the
heat flux.

Formally the initial non-equilibrium state is taken to be generated such that
the time scale for the change in temperature and density is fast compared to
the electron plasma frequency. On the other hand, the time scale for the
generation of these gradients, which happens for example in laser or shock
interactions, is often similar or longer than the period of plasma
oscillations. However, the interaction with the more slowly generated gradients
would only lead to plasma oscillations comparable to those which we have shown
are excited by our initial conditions. In effect, we model the time scale that
is slow compared to the gradient generation, but fast compared to the electron
transit time $L_T/v_{T0}$.  Although as seen above in simulations, even when the
gradients grow on the same order as $L_T/v_{T0}$ ($\delta \propto t$) the
evolution of anisotropy continues to follow \eq{anisotropydef}, with $A \sim
\delta\left(t\right)^2 t^2$.

It should be noted that for simplicity, there are a few limitations to the
generality of this work. The pressure and density gradients are assumed to be
perpendicular, and the gradients are entirely linear, not including second
order variation. The more general case will be reserved for a future work.

However, this solution remains quite general. An anisotropic Maxwellian
distribution ($v_{Ti0} \ne v_{Tj0}$, where $v_{Ti0}$ is the thermal velocity in
the $i$ direction) can be modeled by the same equations. In that case, $\bm x$,
$\bm v$, and $\bm E$ are normalized using the $v_{Ti0}$ in the same direction,
and \eq{Bbiermann} has an additional factor of $v_{Tx0}/v_{Ty0}$. This means
that the Biermann field is caused solely by the thermal spread directed
along the density gradient ($v_{Tx0}$).

Moreover, the kinetic result of anisotropy generation is relevant even for
magnetized cases, as long as there is a temperature gradient parallel to the
magnetic field. Our solution for the case with $\epsilon=0$ is valid for a
uniform parallel magnetic field of arbitrary magnitude.  For significantly
large fields, instabilities driven by the anisotropy in a magnetized plasma,
such as the firehose instability~\cite{Parker58}, would dominate over the
Weibel instability.  

Anisotropy driven instabilities can help explain weak heat fluxes in cooling
flows. Another kinetic instability that can lead to suppression in heat flux is
driven solely by the heat flux~\cite{Roberg2016} with a growth rate
$\gamma_{HF} \approx 0.1 \Omega_{ce} \epsilon_{HF}$, where $\Omega_{ce}$ is the
electron cyclotron time and $\epsilon_{HF}$ is the coefficient proportional to
the heat flux taken from~\cite{Roberg2016}.  The second term on the RHS in
\eq{fTevolution} corresponds to $\epsilon_{HF} = \sqrt{2} \delta t$.  We can
estimate the onset time $\tau_{HF}\approx 2.7 (\delta \Omega_{ce})^{-1/2}$ of
this instability by comparing the predicted heat flux growth $1/\epsilon_{HF}
\partial \epsilon_{HF}/\partial t = 1/t$, to $\gamma_{HF}$.

Comparing the onset time of the heat flux instability to the Weibel $\tau_W
\approx 1.6 (\delta^3 v_T/c)^{-1/4}$ (in the limit $A \ll 1$, where
$\gamma_W(A) = (8/27\pi)^{1/2}A^{3/2}v_T/c$ \cite{Davidson72}), reveals the
Weibel instability will appear first as long as $\beta_e$, the ratio of the
electron plasma pressure to the magnetic pressure, is sufficiently large;
$\beta_e \gtrsim L_T/d_e$. The Biermann battery alone often grows slow enough
that $\beta_e$ remains larger before the Weibel onsets; as long as $\delta
\lesssim 4 ((v_T/c) L_n/L_T )^4$, using $\tau_W$ in \eq{Bbiermann} to find
$\beta_e$. Either of these instabilities is likely to cause the heat flux to
saturate long before reaching the collision time.

The purely kinetic temperature anisotropy generation from temperature
gradients is thus relevant for a wide variety of settings; from astrophysical
shocks and laser experiments with small collision rates where the Biermann
battery can also exist, to flux tubes~\cite{Parker79,Russell90} with
temperature gradients found in the solar corona or at the Earth's magnetopause.

\paragraph{Acknowledgments.}
This work was supported by the European Research Council (ERC-2010-AdG Grant
No. 267841, and ERC-2015-AdG Grant No. 695008).  NFL was partially funded by NSF CAREER award no. 1654168.

%

\end{document}


\setcounter{secnumdepth}{1}

\title{Supplementary Materials for:\\The fully kinetic Biermann battery and associated generation of pressure anisotropy}

\author{K. M. Schoeffler}
\affiliation{GoLP/Instituto de Plasmas e Fus\~ao Nuclear,
Instituto Superior T\'ecnico,\\ Universidade de Lisboa, 1049-001 Lisboa, Portugal}
\author{N. F. Loureiro}
\affiliation{Plasma Science and Fusion Center, Massachusetts Institute of Technology, Cambridge MA 02139, USA}
\author{L. O. Silva}
\affiliation{GoLP/Instituto de Plasmas e Fus\~ao Nuclear,
Instituto Superior T\'ecnico,\\ Universidade de Lisboa, 1049-001 Lisboa, Portugal}

\date{\today}

\begin{abstract}

\end{abstract}


\maketitle


\section{Equations solved.}
The evolution of the electron distribution function subject to these density
and temperature gradients is given by the Vlasov equation, coupled with
Faraday's and Ampere's laws, for quantities $f^{(i)}$, $E^{(i)}$, and
$B^{(i)}$, which are of order $\epsilon^{(i)}$. Initially $f^{(0)} = f_M$, and
$f^{(1)}$ and $f^{(2)}$ are the respective terms on the first and second lines
of Eq.~(3) in the paper.  The initial fields are either 0, or those described
in the paper in Eqs.~(9, 13, 17).

The zeroth order term is trivially solved given $E^{(0)} = B^{(0)} = \nabla f^{(0)} 
= \int d^3v \mathbf{v}f^{(0)}/n_0 = 0$.

For the first order terms:
\begin{equation}
\label{vlasov1}
\frac{\partial f^{(1)}}{\partial t} + \mathbf{v} \cdot \mathbf{\nabla} f^{(1)} 
- \left( \mathbf{E}^{(1)}
+ \mathbf{v} \times \mathbf{B}^{(1)} \right)
\cdot \mathbf{\nabla_{v}} f^{(0)} = 0
\text{,}
\end{equation}
\begin{equation}
\label{faraday1}
\frac{\partial \mathbf{B}^{(1)}}{\partial t} = -\curl \mathbf{E}^{(1)}
\text{,}
\end{equation}
\begin{equation}
\label{ampere1}
\frac{\partial \mathbf{E}^{(1)}}{\partial t} = \int d^3v \mathbf{
v}\frac{f^{(1)}}{n_0} + \frac{c^2}{v_{T0}^2} \curl \mathbf{B}^{(1)}
\text{.}
\end{equation}
We can simplify \eq{vlasov1} noting that $\mathbf{\nabla_{v}} f^{(0)} =
-\mathbf{v}  f^{(0)}$, and therefore the term $\mathbf{v} \times
\mathbf{B}^{(1)}  \cdot \mathbf{\nabla_{v}} f^{(0)} = 0$. 
Furthermore, we can divide $f^{(1)}$ into an odd and even function of $v$
yielding two equations:
\begin{equation}
\label{vlasov1a}
\frac{\partial f_{odd}^{(1)}}{\partial t} + \mathbf{v} \cdot \mathbf{\nabla} f_{even}^{(1)} 
+ \mathbf{E}^{(1)}
\cdot \mathbf{v} f^{(0)} = 0
\text{,}
\end{equation}
\begin{equation}
\label{vlasov1b}
\frac{\partial f_{even}^{(1)}}{\partial t} + \mathbf{v} \cdot \mathbf{\nabla} f_{odd}^{(1)} 
 = 0
\text{.}
\end{equation}

Initially $f^{(1)}$ is even and proportional to $x$ or $y$, and thus the
gradient is uniform in space.  Since the initial $E^{(1)}$ is either 0 or
uniform, \eq{vlasov1a} shows that an odd function is produced, that is also
uniform in space.  Initially $E^{(1)}$, $B^{(1)}$, and $f_{even}^{(1)}$ do not
change since $f^{(1)}$ is even, and $E^{(1)}$ and $B^{(1)}$ are uniform in
space (see \eqsct{faraday1}{ampere1}{vlasov1b}).  Next, the new odd function
generates an electric field seen in \eq{ampere1} which remains uniform in
space.  Moreover, since $f_{even}^{(1)}$ has not changed and the new $E^{(1)}$
is uniform in space, the $f_{odd}^{(1)}$ produced continues to be uniform in
space (see \eq{vlasov1a}).  $B^{(1)}$, and $f_{even}^{(1)}$ continue to not
change since $f_{odd}^{(1)}$ and $E^{(1)}$ are uniform in space (see
\eqsct{faraday1}{ampere1}{vlasov1b}).  At all subsequent times, the same
arguments hold, and $f^{(1)}-f^{(1)}(t=0)$ will continue to be odd in $v$ and
uniform in space.

Note that additional factors of the Lorentz factor
$\gamma=\left(1-v^2/c^2\right)^{-1/2}$ found in the relativistic equations, or
coupling to a similar Vlasov equation for an ion species, by an additional ion
current in \eq{ampere1} proportional to $c_s/v_{T0}$ do not affect these
arguments, such that $f^{(1)}$ will continue to be odd in $v$ and uniform in
space.  (Keep in mind, the ion distribution also evolves according to
\eqs{vlasov1a}{vlasov1b}.) Since the first order terms are all odd in $v$ and uniform in space, we can thus safely conclude that the second order terms
($\sim \epsilon^2$) that are even in $v$ or space dependent (which is in fact
all of them) are only slightly modified by small $c_s/v_{T0}$ and $v_{T0}^2/c^2$
terms $\sim \epsilon^0$.

Although the previous discussion has demonstrated that these small terms can be
neglected, for completeness, we will also perform a similar analysis for the
second order terms. Here we will demonstrate that it is required that none of the
second order solutions are uniform in space unless they are also even with
respect to $v$, or magnetic fields. 
For the second order terms:
\begin{equation}
\label{vlasov2}
\frac{\partial f^{(2)}}{\partial t} + \mathbf{v} \cdot \mathbf{\nabla} f^{(2)} 
- \mathbf{E}^{(1)}
\cdot \mathbf{\nabla_{v}} f^{(1)} 
+ \mathbf{E}^{(2)}
\cdot \mathbf{v} f^{(0)} = 0
\text{,}
\end{equation}
\begin{equation}
\label{faraday2}
\frac{\partial \mathbf{B}^{(2)}}{\partial t} = -\curl \mathbf{E}^{(2)}
\text{,}
\end{equation}
\begin{equation}
\label{ampere2}
\frac{\partial \mathbf{E}^{(2)}}{\partial t} = \int d^3v \mathbf{
v}\frac{f^{(2)}}{n_0} + \frac{c^2}{v_{T0}^2} \curl \mathbf{B}^{(2)}
\text{.}
\end{equation}
Now we can divide $f^{(2)}$ into an odd and even function of $v$:
\begin{equation}
\label{vlasov2a}
\frac{\partial f_{odd}^{(2)}}{\partial t} + \mathbf{v} \cdot \mathbf{\nabla} f_{even}^{(2)} 
- \mathbf{E}^{(1)}\mathbf{\nabla_{v}}f^{(1)}_{even} + \mathbf{E}^{(2)}
\cdot \mathbf{v} f^{(0)} = 0
\text{,}
\end{equation}
\begin{equation}
\label{vlasov2b}
\frac{\partial f_{even}^{(2)}}{\partial t} + \mathbf{v} \cdot \mathbf{\nabla} f_{odd}^{(2)} 
- \mathbf{E}^{(1)}\mathbf{\nabla_{v}}f^{(1)}_{odd} 
 = 0
\text{.}
\end{equation}

Initially $f^{(2)}$ is even and proportional to $x^2$ or $y^2$, and thus the
gradient is proportional to $x$ or $y$.  \eq{vlasov2a} shows that since the
initial $E^{(1)}$ is uniform, and the initial $E^{(2)}$ and $f_{even}^{(1)}$
are either 0 or proportional to $x$ or $y$, an odd function is produced, that
is proportional to $x$ or $y$.  From \eq{ampere2}, $E^{(2)}$ does not change
since $f^{(2)}$ is even, and $B^{(2)}$ is uniform.  From \eq{faraday2},
$B^{(2)}$ grows uniformly in space, because  $E^{(2)}$ is proportional to $x$
or $y$ leading to a uniform gradient.  Likewise $f_{even}^{(2)}$, from
\eq{vlasov2b}, grows uniformly in space, because $f_{odd}^{(1)}$ and $E^{(1)}$
are uniform.  

After the initial stage, the new odd function generates an
electric field seen in \eq{ampere2} which remains proportional to $x$ or $y$,
while the curl of $B^{(2)}$ remains 0.  Moreover, since the additional
$f_{even}^{(2)}$ is uniform and the new $E^{(2)}$ is proportional to $x$ or
$y$, the $f_{odd}^{(2)}$ produced continues to be proportional to $x$ or $y$.
$B^{(2)}$, and $f_{even}^{(2)}$ continue to grow uniformly since
$f_{odd}^{(2)}$ and $E^{(2)}$ are proportional to $x$ or $y$, and
$f_{odd}^{(1)}$ and $E^{(1)}$ remain uniform.  At all subsequent times, the
same arguments hold, and the $E^{(2)}$ and $f_{odd}^{(2)}$ produced are all
proportional to $x$ or $y$, while the $B^{(2)}$ and $f_{even}^{(2)}$ produced
are all uniform in space.  Again this analysis was for the sake of
completeness, but it further justifiies neglecting the small terms
$c_s/v_{T0}$ and $v_{T0}/c$.

In order to find the explicit time dependence of $E^{(1)}$, $E^{(2)}$,
$B^{(2)}$, $f_{odd}^{(1)}$, $f_{odd}^{(2)}$, and $f_{even}^{(2)}$, we assume
the initial condition of $f = f_0$, no initial electric or magnetic fields (or
guess an equilibrium field), and take an expansion for small $t\omega_{pe}$.
For example we can use \eq{faraday1} to find $f_{odd}^{(1)}$ to first order in
$t\omega_{pe}$, then \eq{ampere1} to find $E^{(1)}$ and \eq{faraday1} to find
$f_{odd}^{(1)}$ to second order in $t\omega_{pe}$, and contine to higher orders
in $t\omega_{pe}$.  Fortunately, summing over all orders of $t$ converges to an
analytic solution valid for $t \sim \epsilon^0$.  The same procedure is used to
calculate the second order terms in $\epsilon$.

\section{Density gradient.}
We first consider the case with only a density gradient ($\delta=0$). If we
assume the initial condition of $f = f_0$ and no initial electric or magnetic
fields, the solution to the Vlasov Maxwell's equation was described
as:
\begin{equation}
\label{perturbedfdensity}
f = f_0 + \tilde{f_n} \left(t\right)
\text{.}
\end{equation}
Here the oscillatory term is defined as:
\begin{multline}
\label{perturbedfdensity}
\tilde{f_n} \left(t\right) \equiv  
-\epsilon  \sin\left(\omega_{pe,x}t\right) v_x f_M \\
+\frac{1}{2}
\epsilon^2
 \sin\left(\omega_{pe,x}t\right) x  v_x 
 f_M\\
-\epsilon^2 \left[1-\cos\left(\omega_{pe,x}t\right)
\right]
v_x^2 
f_M\\
+\frac{1}{2}\epsilon^2 t \sin\left(\omega_{pe,x}t\right)
v_x^2 
f_M\\
-\frac{1}{2}\epsilon^2  \left[1-\cos\left(\omega_{pe,x}t \right)\right]^2
\left(v_x^2 -1\right) 
f_M
\text{,}
\end{multline}
where $\omega_{pe,x} \equiv 1+\epsilon x/2$ is the
normalized plasma frequency based on the $x$ dependent density, $n$.  Note that we
have made use of the Poincar\'e-Lindstedt method
\cite{Lindstedt1882,Poincare1893}, which by modifying the frequency in the
solution, avoids unphysical secularly growing terms. This is done by including
additional higher order terms ($> \epsilon^2$) found in the expansion of the
sin and cos terms. 

The third and last terms on the RHS of \eq{perturbedfdensity} are associated
with a modification to the density, temperature, and temperature anisotropy of
the plasma. Although a component of each of these terms does not oscillate with
time, these terms are insignificant because this modification is of order
$\epsilon^2$, and does not grow with time.

Despite the use of the Poincar\'e-Lindstedt method
\cite{Lindstedt1882,Poincare1893},  which avoids unphysical secularly growing
terms, there still exists a secular term in $f$, which grows linearly with
time until $t \sim \epsilon^{-1}$ where the
assumptions of the ordering break. Thus our model is only valid as long as $t$
remains small compared to this limit. This term is, however, physical, and
represents the increasing electron density associated with the divergence of
the electric field: 
\begin{multline}
\bm \nabla \cdot \mathbf{E} = (n_i-n_e)/n_0\\ =
\epsilon^2 
 \left[1 - \cos\left(\omega_{pe,x}t\right) - \frac{1}{2}\omega_{pe}t \sin\left(\omega_{pe,x}t\right) \right]
\text{.}
\end{multline}
This increasingly large divergence is caused by the space dependent frequency,
$\omega_{pe,x}$, which gives rise to increasingly shorter scale variations
along $x$ in the electric field.  These variations along $x$ lead to phase
mixing and Landau damping.  At late time, where the assumptions break down,
Landau damping eventually eliminates both the oscillations and the secular
term.

In the case where there in an initial electric field such that there are no
oscillatory terms,
\begin{equation}
\bm \nabla \cdot \mathbf{E} = \epsilon^2 \text{,}
\end{equation}
so there is only a slight constant and uniform difference between $n_i$ and
$n_e$. 

\begin{figure*}[!t]
  \noindent\includegraphics[width=6.0in]{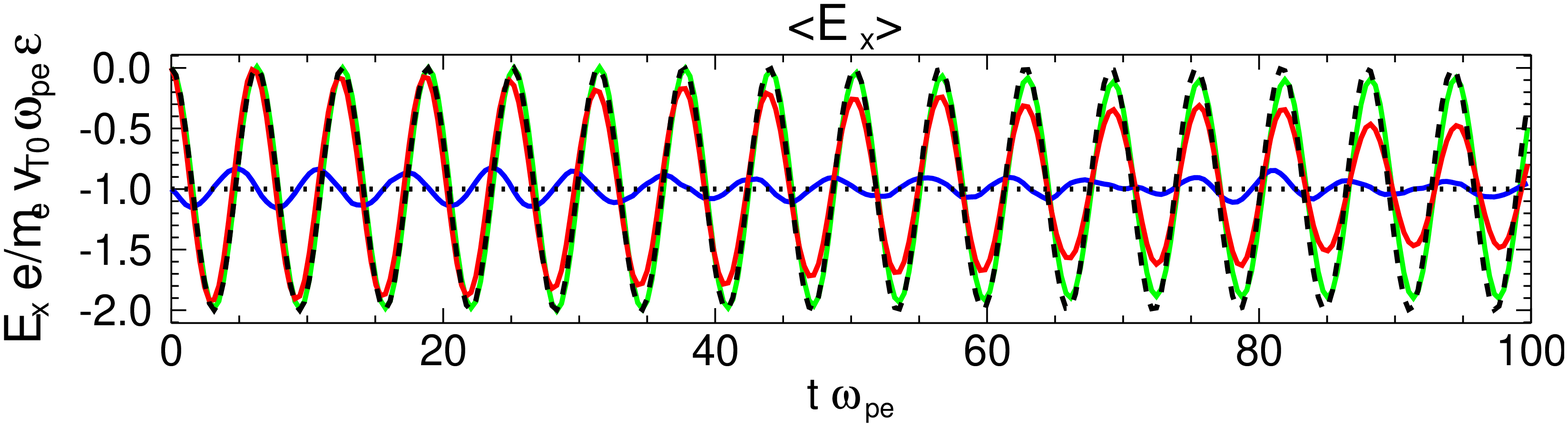}
  \caption{\label{Exoscillation}
  Average $E_x$ vs. time due to the density gradient starting from $0$ electric
  field (red), and starting from \eq{equilibriumEdensity} (blue), along with the
  predicted curves from \eqsc{perturbedEdensity}{equilibriumEdensity} (black
  dashed/dotted lines respectively).  Average $E_x$ starting from $\bm E = 0$
  between $x/\lambda_D=-10, 10$, and $y/\lambda_D=-490, 490$, with 64,000 ppg
  (green) instead of 8,000 ppg.
}
\end{figure*}
\begin{figure*}[!t]
  \noindent\includegraphics[width=6.0in]{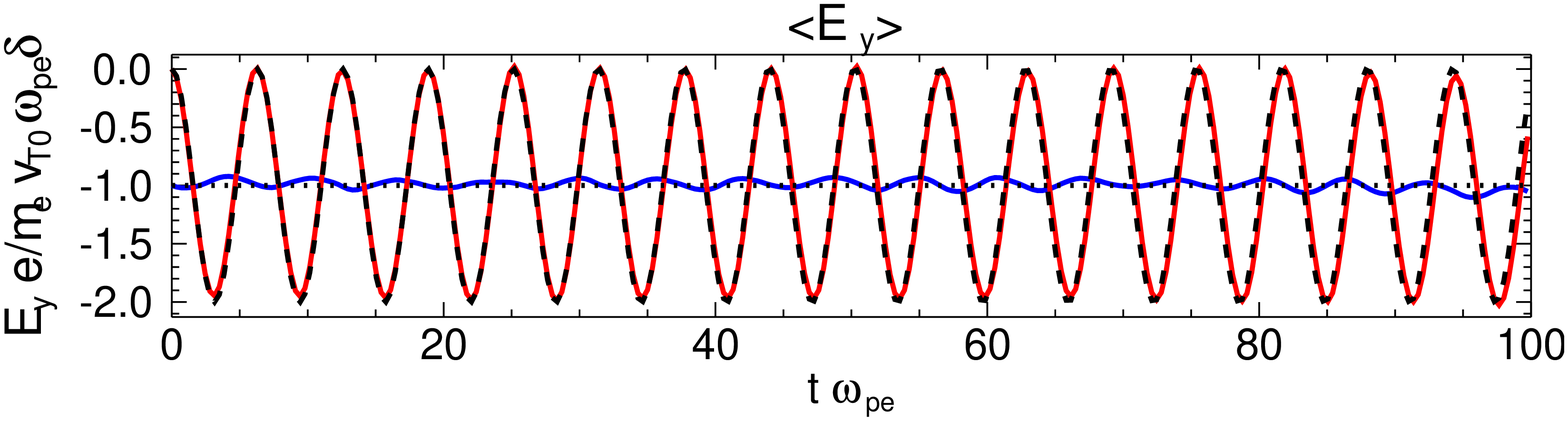}
  \caption{\label{Eyoscillation}
 Average $E_y$ vs. time due to the temperature gradient starting from $0$ electric field (red), and starting from \eq{equilibriumEpressure} (blue), 
along with the predicted curves from \eqsc{perturbedEpressure}{equilibriumEpressure} (black dashed/dotted lines respectively). 
}
\end{figure*}
\FloatBarrier

\section{Temperature gradient.}
We now consider the second case, with only a temperature gradient ($\epsilon=0$). If we
again assume the initial condition of $f = f_0$ and no initial electric or magnetic
fields, the solution to the Vlasov Maxwell's equation was described
as:
\begin{equation}
\label{perturbedfpressure}
f = f_{\nabla T} + \tilde{f_T}\left(t\right)
\text{.}
\end{equation}
Here the oscillatory term is defined as:
\begin{multline}
\label{perturbedfpressure}
\tilde{f_T}\left(t\right) \equiv
- \delta \sin\left(t \right)v_y f_M\\
+\frac{1}{2}\delta^2 \sin\left(t \right)y \left(5-v^2\right) v_y f_M\\
-\frac{1}{2}\delta^2 \left[1-\cos\left(t \right)\right]\left(5-v^2\right)v_y^2 f_M\\
+\frac{1}{2}\delta^2 \left[1-\cos\left(t \right)\right]\left[v_y^2\left(7-v^2\right) -\left(5 - v^2\right)\right] f_M \\
-\frac{1}{2}\delta^2 t \sin\left(t \right)\left[v_y^2\left(7-v^2\right) -\left(5 - v^2\right)\right] f_M \\
-\frac{1}{2}\delta^2 \left[1-\cos\left(t \right)\right]^2\left(v_y^2-1\right) f_M
\text{.}
\end{multline}

Like the previous case, there is a term proportional to $t$ in the distribution
function $\tilde{f_T}\left(t\right)$, which eventually breaks the assumptions
of the ordering. However, while the assumptions hold, this term is not so
significant as it does not contribute to either the temperature, or the density
of the system.

In addition, the third and last terms on the RHS of \eq{perturbedfpressure} are
associated with a modification to the density, temperature, and
temperature anisotropy of the plasma, that does not oscillate with time.
However, these terms are also insignificant because this modification is of
order $\delta^2$, and does not grow with time.

In a system with a density gradient, both these oscillations, and those found
in the electric field would Landau damp as described earlier.

\section{Numerical comparison.}
Our theoretical model has been tested via particle-in-cell (PIC) simulations
using the OSIRIS framework \cite{Fonseca02, Fonseca08}, and we have obtained
agreement with the predicted anisotropic heating and magnetic field generation.
Here we will explain our initial setup and parameters, as well as show further
agreement with our model.

Unless otherwise specified, the simulations are performed with the parameters
$L_T/\lambda_D = L_n/\lambda_D = 1000 = \epsilon^{-1}$, and a box half width of
$L_x/\lambda_D = 49$, and $L_y/\lambda_D = 490$. The normalized thermal
velocity has been chosen small, $v_{T0}/c = 0.05$, such that relativistic
effects do not play a role, but it is large compared to $\epsilon=0.001$. At
the ends of the density gradient the particles have reflecting boundary
conditions, and at the ends of thermal gradient there is a thermal bath. On all
sides there are magnetic conducting boundary conditions.  The box is longer in
the $y$ direction to isolate the central region of the simulation from possible
non-physical boundary effects due to the thermal bath.  Our simulations run
until $\omega_{pe}t=100$, while $\omega_{pe}L_y /v_{T0} = 490$.  The resolution
is $1.4$ grid points per Debye length ($\lambda_D$) and the time resolution is
$\Delta t \omega_{pe} = 0.0175$.  In order to resolve the small order details,
we use a large number of particles per grid cell (ppg) of $8,000$. All of the
presented simulations are done with non-moving ions like assumed in the
calculation.  

To test these solutions we look at both simulations with $\delta=0$ or
$\epsilon=0$, (periodic boundary conditions in the direction with no gradient),
and with both gradients.  

Here we further test the model by examining the equilibrium static electric
fields in the cases with $\delta=0$ predicted to be:
\be
\label{equilibriumEdensity}
\bm{E} = -
\left(\epsilon
-\epsilon^2
x  \right)
\hat{\bm x}
\text{,}
\ee
and for $\epsilon=0$: 
\be
\label{equilibriumEpressure}
\bm{E}
= - \delta
\hat{\bm y}
\text{,}
\ee
and for oscillatory electric fields generated when the initial electric field is zero, in the cases with $\delta=0$:
\begin{equation}
\label{perturbedEdensity}
\bm{E} 
= -
\left(\epsilon
-\epsilon^2
x \right)
\left[1-\cos\left(\omega_{pe,x}t\right)\right]
\hat{\bm x}
\text{,}
\end{equation}
and for $\epsilon=0$: 
\be
\label{perturbedEpressure}
\bm{E}
=
-\delta
\left[1-\cos\left(t \right)\right]
\hat{\bm y}
\text{.}
\ee

In~\fig{Exoscillation}, one can see the oscillations in the average electric
field from the simulation with $\delta=0$ following \eq{perturbedEdensity}. The
decay in amplitude is an effect of the averaging with
$\omega_{pe,x}$, which is a function of $x$, and so a simulation with a smaller range in
$x/\lambda_D$ (-10,10) from a simulation with better statistics (64,000 ppg) is
also shown for comparison. In a simulation starting with
\eq{equilibriumEdensity}, the electric field stays at the equilibrium value.
Similarly, in~\fig{Eyoscillation} the oscillations in the electric field from
the simulation with $\epsilon=0$ follow \eq{perturbedEpressure}, and a
simulation starting with \eq{equilibriumEpressure}, shows the electric field
staying at the equilibrium value.  

In the case with both gradients, since the boundary conditions significantly
affect the magnetic fields close to the boundaries (at the boundary $B_z=0$),
it was necessary to further increase the box size by a factor of $10$ in the
$y$ direction.  The ordering assumptions are only met where $\bm{x}/\lambda_D
\sim \epsilon^{0}$, and a simulation where this is true for a significant range
would be unfeasible. The fields are calculated by averaging the results between
$-49< x/\lambda_D <49$, and $-49 < y/\lambda_D < 49$.  For the simulation with
both $\delta \ne 0$ and $\epsilon \ne 0$, similar curves with constant electric
fields as shown in blue in~\figs{Exoscillation}{Eyoscillation} were obtained.

\section{Acknowledgments.}
This work was supported by the European Research Council (ERC-2010-AdG Grant
No. 267841, and ERC-2015-AdG Grant No. 695008).  NFL was partially funded by NSF CAREER award no. 1654168.

%